\documentclass[onecolumn,12pt,journal,draftclsnofoot,journalpaper]{IEEEtran}

\usepackage{amssymb}
\usepackage{amsmath}
\usepackage{amsthm}
\usepackage{amsfonts}
\usepackage{mathrsfs}
\usepackage{float}
\usepackage{graphicx}
\usepackage{epsfig,graphics,subfigure,psfrag}
\usepackage{pdfpages}
\usepackage{enumerate}
\usepackage{engord}
\usepackage{cite}
\usepackage{verbatim}
\usepackage[ruled]{algorithm2e}

\begin{document}

\newtheorem{definition}{\bf Definition}
\newtheorem{theorem}{\bf Theorem}
\newtheorem{lemma}{\bf Lamma}
\newtheorem{proposition}{\bf Proposition}

\title{Caching as a Service: Small-cell Caching Mechanism Design for Service Providers}
\author{
\IEEEauthorblockN{
\normalsize{Zhiwen Hu},
\normalsize{Zijie Zheng},
\normalsize{Tao Wang},
\normalsize{Lingyang Song},
and
\normalsize{Xiaoming Li}
 \\}
\IEEEauthorblockA{\normalsize{School of Electronics Engineering and Computer Science\\ Peking University, Beijing, China\\
Email: \{zhiwen.hu, zijie.zheng, wangtao, lingyang.song, lxm\}@pku.edu.cn} \\
}
\thanks{Part of the material in this paper was accepted by IEEE International Conference on Computer Communications (INFOCOM15), Hong Kong, Apr.~2015 \cite{bib_INFOCOM}. }
}
\maketitle

\begin{abstract}

Wireless network virtualization has been well recognized as a way to improve the flexibility of wireless networks by decoupling the functionality of the system and implementing infrastructure and spectrum as services.
Recent studies have shown that caching provides a better performance to serve the content requests from mobile users.
In this paper, we propose that \emph{caching can be applied as a service} in mobile networks, i.e., different service providers (SPs) cache their contents in the storages of wireless facilities that owned by mobile network operators (MNOs).
Specifically, we focus on the scenario of \emph{small-cell networks}, where cache-enabled small-cell base stations (SBSs) are the facilities to cache contents.
To deal with the competition for storages among multiple SPs, we design a mechanism based on multi-object auctions, where the time-dependent feature of system parameters and the frequency of content replacement are both taken into account.
Simulation results show that our solution leads to a satisfactory outcome.

\end{abstract}


\newpage
\section{Introduction}

\emph{Wireless network virtualization} has been proposed in recent years to improve the flexibility of traditional wireless networks against the tremendous growth of diversified online services~\cite{bib_SurveyOfWNV1}.
Similar to the traditional wired network virtualization \cite{bib_SurveyOfNV1}, wireless network virtualization separates wireless networks into physical infrastructures and online services \cite{bib_VirtualResources}.
In wireless networks, the parties that operate the wireless physical infrastructures are called mobile network operators~(MNOs), and the parties that provide online services for users are called service providers~(SPs).
SPs can typically create their own virtual networks to serve their users by aggregating resources from MNOs, where the \emph{resources} usually have a broad scope, ranging from the spectrum, the infrastructure, to the air interface \cite{bib_SurveyOfWNVRA}.
With the help of virtualization, multiple heterogeneous virtual networks that dynamically composed by different SPs can coexist together in isolation from each other \cite{bib_VirtualResources}.
Therefore, once the system is properly designed, wireless network virtualization can maximize the system utilization, facilitate the updating of existed services and alleviate the difficulty of applying new ones \cite{bib_SurveyOfWNV2}.

Since the services provided by SPs depend on the resources that allocated to them, resource allocation becomes one of the important issues, i.e. how to effectively allocate the limited resources to different SPs \cite{bib_SurveyOfWNV1}.
In most early studies, spectrum was considered as the most basic kind of resource in wireless network virtualization.
The authors in \cite{bib_Allocation1}\cite{bib_Allocation2} discussed the spectrum allocation problem in both time domain and frequency domain, and the works in \cite{bib_Allocation3}\cite{bib_Allocation4} dealt with the competition for spectrum among SPs by using game theory.
Apart from the spectrum, another kind of important resource that being considered in previous works was the infrastructure, such as the wireless building premises, RF antennas, and network routers, etc \cite{bib_SurveyOfWNV1}.
Several studies showed the ongoing trends of the virtualization of wireless infrastructures \cite{bib_Infrastructure1}\cite{bib_Infrastructure2}.
Moreover, the combination of spectrum and infrastructure sharing was proposed as \emph{full network sharing}, which was detailedly classified in \cite{bib_Fullsharing}.

However, there are still potential resources that are not discussed in wireless network virtualization, such as the storages of wireless facilities\cite{bib_CachingInTheAir}.
The storage-enabled wireless facilities were proposed in \cite{bib_RelatedFemtoCaching,bib_RelatedApproximation,bib_RelatedMulticast,bib_CachingComplexity,bib_RelatedCoded1,bib_RelatedCoded2,bib_RelatedSocial,bib_RelatedMobility}, where the pre-cached contents in small-cell base stations (SBSs) can bring better system performance, showing an effective way to deal with the low-speed backhaul link of SBSs \cite{bib_SmallCell1}.
This proposal was first given in \cite{bib_RelatedFemtoCaching}, where a sub-optimal strategy of caching content was provided.
Based on this, authors of \cite{bib_RelatedApproximation,bib_RelatedMulticast,bib_CachingComplexity} considered a more detailed physical layer model.
Other studies in \cite{bib_RelatedCoded1}\cite{bib_RelatedCoded2} further discussed the network layer coding technique.
Moreover, the works in \cite{bib_RelatedSocial}\cite{bib_RelatedMobility} also took into account the social ties and the mobilities of users respectively.

Although small-cell caching has been discussed from many aspects, few existing studies focus on the decision layer, where multiple SPs may exist.
Since all the SPs only intend to better serve their own users by caching their own contents to reduce the average delay, they are likely to compete for limited caching storages.
Thus a proper mechanism should be designed to deal with the competition among SPs and guarantee the overall performance at the same time.
To solve the aforementioned problem, we propose to design an effective caching mechanism, which enables \emph{caching as a service} in wireless networks.
Specifically, the storages of wireless facilities can be virtualized and shared among multiple SPs, and these SPs can utilize the storages as caching spaces to cache their own contents for their users.
With the help of caching, the average delay of content requests from users can be lowered, such that the quality-of-service can be improved.
Here we have to clarify the differences between our work and the studies in \cite{bib_ChannelAuction1}, \cite{bib_ChannelAuction2} and \cite{bib_CachingAsAService}, where the authors of \cite{bib_ChannelAuction1} and \cite{bib_ChannelAuction2} take wireless channels instead of caching storages as the objects to be auctioned, and the authors of \cite{bib_CachingAsAService} propose that caching services are provided by the MNO who cache the contents in a centralized way instead of by multiple SPs who have to compete with each other for their own users.

Without loss of generality, in this paper, we focus on a scenario where small-cell base stations are the facilities that used to cache contents \cite{bib_RelatedFemtoCaching}.
We formulate the multiple SPs' small-cell caching problem by taking into account the overlapping among SBSs.
Since SPs have to compete for the caching storages on behalf of their own contents, a nature solution is to apply auctions \cite{bib_AuctionTheory}, where each SP has to evaluate its contents and bid for caching storages.
We propose our own mechanism based on multi-object auctions, where the mechanism organizes a serial of multi-object auctions to complete the caching scheme.
Each multi-object auction can be solved by the market matching algorithm \cite{bib_MultipleItems}, which takes valuations as input and takes allocation results and prices as output.
Considering that the system parameters are time-dependent, storages of SBSs may change contents to adapt to the variation, which also burdens the backhaul link of SBSs.
To cope with this problem, we also present a novel approach to reduce the frequency of content replacement.
Simulation results have shown the effectiveness of our solution.

The main contributions of our work are listed below:
\begin{enumerate}
\item We come up with a novel approach that caching can be applied as a service in the mobile networks with the help of wireless network virtualization, where each SP has to pay for the storages of the infrastructure that owned by MNOs.
\item We focus on the small-cell caching scenario and formulate the caching problem with the objective to minimize average delay, where the overlapping among SBSs and the competition among SPs are considered.
\item By using multi-object auctions in our mechanism, we provide a sub-optimal solution and also find a way to reduce the frequency of content replacement between adjacent hours.
\end{enumerate}

The rest of our paper is organized as follows.
Section II presents our system model of small-cell caching.
Section III provides the problem formulation and the theoretical analysis on the system parameters.
Section IV introduces our auction mechanism.
Section V shows the simulation results which prove the effectiveness of our solution and testify our theoretical analysis.
Finally, we conclude our paper in Section VI.

\section{System Model}

\begin{figure}[!bht]
\centering
\includegraphics[width=3.5in]{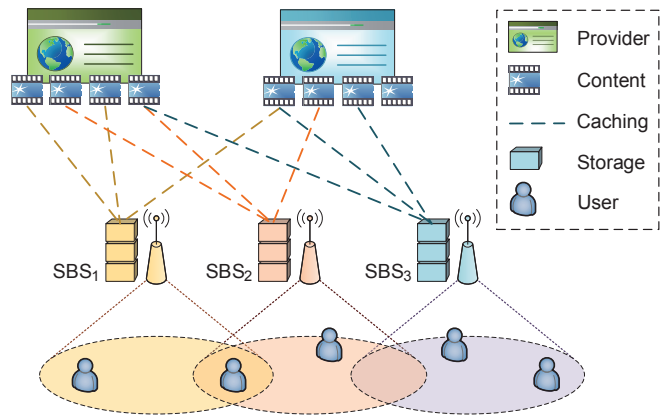}
\vspace{-6mm}
\caption{System model for small-cell caching among multiple providers. For each request from any user, the delay can be lowered as long as the requested content is cached in a nearby SBS.}\label{fig_SystemModel}
\vspace{-6mm}
\end{figure}

In this paper, we study a small-cell network, which involves $I$ SBSs in an area and $L$ SPs that provide different contents for users, as shown in Fig.~\ref{fig_SystemModel}.
We use SBS$_i$ to denote the $i^{th}$ SBS, and SP$_l$ to denote the $l^{th}$ service provider, where $1 \le i \le I$ and $1 \le l \le L$.
These SPs intend to cache their own contents into SBSs, where the storage capacity of SBS$_i$ is given by $H_i$.

For the rest part of this section, we model our system in four aspects: the coverage region of SBSs, the distribution of users, the contents of SPs, and the traffic latency of content requests.

\begin{figure}[!thp]
\centering
\includegraphics[width=2.2in]{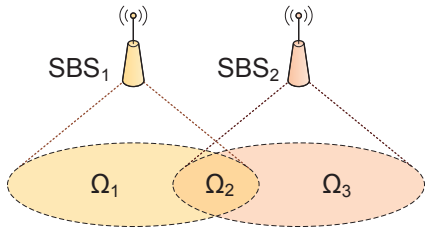}
\vspace{-6mm}
\caption{A demonstration of coverage regions of two overlapping SBSs, where three simplest regions are included.}\label{fig_CoverageRegion}
\vspace{-6mm}
\end{figure}

\textbf{Coverage regions}:
We assume that the SBSs are distributed in a 2-dimensional area, and the coverage regions of SBSs overlap with each other, as shown in Fig.~\ref{fig_CoverageRegion}.
We define a region as a \emph{simplest region} if it is not crossed by any curves in such a figure.
In our model, simplest regions are denoted by $\Omega_j$, $1 \le j \le J$, where $J$ is the total number of simplest regions.
Since $\Omega_j$ is covered by a set of SBSs, we use $\mathcal{F}_j$ to denote the set of SBSs that cover $\Omega_j$.
For the example shown in Fig.~\ref{fig_CoverageRegion}, we have $\mathcal{F}_1=\{1\}, \mathcal{F}_2=\{1,2\}, \mathcal{F}_3=\{2\}$.

\vspace{-6mm}

\textbf{Users' distribution}:
The distribution of mobile users is described by user density, which can be space-dependent as well as time-dependent.
And it can be estimated by some statistical methods \cite{bib_LearingMechanism} with the help of the data collected by SBSs.
In order to better reflect the time-dependent characteristics, we use $t$ to represent a specific time slot, where $t \ge 1$.
And without loss of generality, we assume the length of a time slot is an hour\footnote{One hour's length is a reasonable choice for caching replacement, because one hour can be precise enough to describe the overall variation of user density and content popularity. Although shorter time slot might be a better choice, the key point of our paper is not to choose the best time slot but to solve the problem with given length of time slot.}.
We use $u^t(\boldsymbol x)$ to denote the average user density at the $t^{th}$ hour at location $\boldsymbol x$, where $\boldsymbol x$ is a two dimensional vector in space.
Thus, the average user number at the $t^{th}$ hour in the region $\Omega_j$ can be given by
\begin{equation}\label{eqn_User}
U^t_j=\iint_{\Omega_j} u^t(\boldsymbol x)  d \boldsymbol x.
\end{equation}
The average user number at the $t^{th}$ hour under SBS$_i$ is
\begin{equation}
U^{t,i}=\sum_{j | i \in \mathcal{F}_j} \iint_{\Omega_j} u^t(\boldsymbol x)  d \boldsymbol x.
\end{equation}
And similarly, the total user number at the $t^{th}$ hour can be calculated as:
\begin{equation}\label{eqn_SumUser}
U^t_{sum}=\sum_{j=1}^J \iint_{\Omega_j} u^t(\boldsymbol x)  d \boldsymbol x.
\end{equation}

\textbf{Contents of SPs}:
We assume that SPs possess different sets of contents, and the contents may have different sizes.
The $k^{th}$ content of SP$_l$ is denoted by $C_{l,k}$, and the size of  $C_{l,k}$ is denoted by $S_{l,k}$, where $1 \le k \le K_l$ and $K_l$ is the number of contents possessed by SP$_l$.
At the $t^{th}$ hour, the possibility of $C_{l,k}$ being requested by each single user is described by its \emph{popularity}, denoted by $\phi^t_{l,k}$.
And we also have
\begin{equation}\label{eqn_SumPopularity}
\phi^t_{sum} = \sum_{l,k} \phi^t_{l,k},
\end{equation}
where $\phi^t_{sum}$ is not necessary to be normalized to one, since each user can request several contents in an hour.
A greater $\phi^t_{sum}$ implies more requests from users in an hour.
Note that the trend of the variation of content popularity can also be predicted with some learning mechanisms \cite{bib_PopularityLearning}.

Since contents can be divided into sequential blocks for caching, we use $C_{l,k,n}$ to denote the $n^{th}$ block of $C_{l,k}$, and $S_{l,k,n}$ to denote the size of it,  where $1 \le n \le N_{l,k}$, and $N_{l,k}$ is the number of blocks that $C_{l,k}$ is divided.
Note that this content division procedure does not change users' requesting probability, therefore, all the blocks from the same original content $C_{l,k}$ have the same possibility to be requested, i.e., they share the same popularity $\phi^t_{l,k}$.
Also notice that if one block is being requested, other blocks from the same content are also being requested at the same time, thus their requesting possibility is inter-dependent.

\textbf{Traffic latency}:
If the content requested by a certain user is cached in one of his nearby SBSs, then the request can be served by this SBS, which leads to a lower delay.
Otherwise, one of the nearby SBSs can serve the user by setting up backhaul connections to the core network and downloading the content from the server.
Therefore, the delay model of backhaul-link (from SBSs to SPs) and the delay model of downlink (from SBSs to users) should be constructed.

Here, we assume that the delay of backhaul-link $\theta^t_{back}$ is proportional to $U^t_{sum}$, since the load of the backhaul network and the core network mainly depends on the total number of connected users.
And for SBS$_i$, we assume that the delay of its downlink $\theta^{t,i}_{down}$ is proportional to $U^{t,i}$, i.e., the number of user that SBS$_i$ covers\footnote{Based on the result of \cite{bib_DelayProfile}, the transmission delay is mainly influenced by the number of connected users, and this relationship can be approximately regarded as a linear one. On the other hand, a specific user in the overlapping area of two SBSs can contribute incremental delay to both SBSs, no matter which SBS it is connected to. This is because SBSs may use the same bandwidth and this user takes up a specific channel of both SBSs. Therefore, $\theta^{t,i}_{down}=\beta_2 \cdot U^{t,i}$ is a reasonable assumption.}.
So we have
\begin{equation}\label{eqn_DelayBackhaul}
\theta^t_{back}=\beta_1 \cdot U^t_{sum},
\end{equation}
\begin{equation}\label{eqn_DelayDownlink}
\theta^{t,i}_{down}=\beta_2 \cdot U^{t,i}.
\end{equation}

For a specific user that covered by several SBSs, it will choose a SBS with the lowest delay to download its desired content (since the content may only be cached in a few of these SBSs).
In the ``choosing" procedure, additional delay may be induced, and we assume this kind of delay depends on the number of available SBSs for the user.
And we give the following definition:
\begin{equation}\label{eqn_DelayChoose}
\theta^{j}_{choose}=\beta_3 \cdot |\mathcal{F}_j|,
\end{equation}
where $|\mathcal{F}_j|$ is the number of SBSs by which $\Omega_j$ is covered.
The more SBSs cover a user, the more time will be spent to choose the best downloading SBS.
And we call $\theta^t_{choose}$ as ``choosing delay" later in our paper.

\section{Problem Formulation and Analysis}\label{sec_Theoretical}

In this section, we first formulate the problem and provide the objective function, then analyse the impact of three system parameters, which are the total number of the contents, the average storage capacity of SBSs, and the overlapping percentage.

\vspace{-5mm}
\subsection{Problem Formulation}

We first use $\Gamma$ to denote the allocation matrix, the definition of its elements is given below:
\begin{equation}\label{eqn_Allocation}
\gamma_{l,k,n}^{t,i} =  \left\{
            \begin{array}{lcl}
           1, & &  \textrm{if $C_{l,k,n}$ is cached in SBS$_i$ at $t$}, \\
            0,& & \textrm{if $C_{l,k,n}$ is not cached in SBS$_i$ at $t$}.
            \end{array}
            \right.
\end{equation}

For a user in the region $\Omega_j$, if he requests $C_{l,k}$ at $t$, the delay can be calculated as:
\begin{equation}\label{eqn_SpecificDelay}
\theta^{t,j}_{l,k} = \sum\limits^{N_{l,k}}_{n=1} \dfrac{S_{l,k,n}}{S_{l,k}}  \min\limits_{i|i \in \mathcal{F}_j} \left[ \theta^{t,i}_{down}+(1-\gamma_{l,k,n}^{t,i})\theta^t_{back} \right] +\theta^j_{choose},
\end{equation}
where $\dfrac{S_{l,k,n}}{S_{l,k}}$ is the weight of the $n^{th}$ block of $C_{l,k}$, and the average delay of requesting a specific content should be the weighted summation of the delay of requesting its blocks (which conforms to user experience).

Our main objective is to minimize the average delay of content requests from users at each hour by properly designing the allocation of caching storages.
Based on (\ref{eqn_User}), (\ref{eqn_SumUser}), (\ref{eqn_SumPopularity}) and (\ref{eqn_SpecificDelay}), the average delay at the $t^{th}$ hour can be written as:
\begin{equation}\label{eqn_AverageDelay}
D(t) =   \dfrac{1}{\phi^t_{sum}U^t_{sum}} \sum\limits_{l,k,j} \theta^{t,j}_{l,k} \cdot U^t_j \cdot \phi^t_{l,k}.
\end{equation}

Finally, we give the objective function and its constraint as:
\begin{equation}\label{eqn_Objective}
\min\limits_{\Gamma}\dfrac{1}{\phi^t_{sum}U^t_{sum}} \sum\limits_{l,k,j} \theta^{t,j}_{l,k} \cdot U^t_j \cdot \phi^t_{l,k},  \quad\quad \forall t,
\end{equation}
\begin{equation}\label{eqn_Constraint}
s.t.~\sum_{l,k,n}\gamma_{l,k,n}^{t,i} \cdot S_{l,k,n} \leq H_i,   \quad\quad\quad \forall i, \forall t.
\end{equation}

This problem is hard to optimize, even a much simpler version of this problem given in \cite{bib_RelatedFemtoCaching} is also proved to be NP-hard by reducing to a k-Disjoint Set Cover Problem \cite{bib_ProofOfNP}.
The mechanism given in Section III is a sub-optimal solution based on optimizing a sequence of sub-problems.

\vspace{-5mm}
\subsection{Theoretical Analysis of System Parameters}

In this subsection, we analyse some of the parameters which can affect the performance of the system at each certain hour.
Since the competition for limited caching storages among SPs is the core issue, the total number of the contents to be cached and the storage capacity of SBSs are the two most concerns.
Besides, the degree of overlapping among SBSs can also affect the outcome, which was never quantitatively discussed in early works.
Therefore we analyse three parameters here: the total number of the contents, denoted by $K$; the average storage capacity of SBSs, denoted by $H$; and the overlapping percentage, denoted by $O$.
We define them as:
\vspace{-1mm}
\begin{equation}
\vspace{-1mm}
K =   \sum\limits_{l=1}^L K_l  \,,
\end{equation}
\begin{equation}
\vspace{-2mm}
H =  \sum\limits_{i=1}^I H_i \Big/ I \,,
\end{equation}
\begin{equation}
\vspace{-2mm}
O =  \Big[  \sum\limits_{i=1}^I A_i - A_{total}  \Big]  \Big/ A_{total}  \,,
\end{equation}
where $A_i$ is the area of the coverage region of SBS$_i$, and $A_{total}$ is the total area of SBSs' coverage regions.
Due to the overlap of SBSs, we have $\sum_{i=1}^I A_i \ge A_{total}$, which means $O \ge 0$.
The system performance is mainly reflected and measured by the average delay given in (\ref{eqn_AverageDelay}).
Here, we provide three propositions on the influence of these parameters and proof them respectively.

\vspace{-3mm}
\begin{proposition}
With a certain distribution of user density and content popularity, the total number of the contents $K$ has a positive correlation with the average delay $D(t)$.
\end{proposition}
\vspace{-3mm}
\begin{IEEEproof}
Suppose that there are initially $K$ contents in the system and we denote the set of these content as $\mathcal{C}$.
The average delay can be calculated after allocation, and we have
\begin{equation}\label{eqn_Proposition1-1}
D(t) =   \dfrac{1}{\phi^t_{sum}U^t_{sum}} \sum\limits_{l,k,j} \theta^{t,j}_{l,k} \cdot U^t_j \cdot \phi^t_{l,k} = \dfrac{1}{\phi^t_{sum}U^t_{sum}} \cdot D(t,\mathcal{C}),
\end{equation}
where $D(t,\mathcal{C})$ represents the un-normalized total delay of requesting contents in $\mathcal{C}$.

When additional set of contents $\mathcal{C}^\prime$ with the same popularity distribution is added, supposing $|\mathcal{C}^\prime|=K^\prime = x K$ and $x>0$, we can provide $\phi^{t\,\,\,\,\,\,\prime}_{sum} = (1+x) \phi^t_{sum}$, because the distribution of content popularity are fixed.
The new caching result makes the average delay change in the form as below:
\begin{equation}\label{eqn_Proposition1-3}
D^{\,\prime}(t) = \dfrac{1}{\phi^{t\,\,\,\,\,\,\prime}_{sum}U^t_{sum}} \cdot \left[ D^{\prime}(t,\mathcal{C}) + D^{\prime}(t,\mathcal{C^{\prime}}) \right]
\end{equation}
where the $D^{\prime}(t,\mathcal{C}) $ represents the un-normalized total delay of requesting original contents, and $D^{\prime}(t,\mathcal{C^{\prime}}) $ represents the un-normalized total delay of requesting newly added contents.

Due to the competition brought by additional contents, some of the original contents are evicted from the caching storage, which leads to $D(t,\mathcal{C}) >  D^{\prime}(t,\mathcal{C})$.
And due to the same popularity distribution of $\mathcal{C}$ and $\mathcal{C}^\prime$, the proportion that contents from $\mathcal{C}$ are cached and the proportion that contents from $\mathcal{C}^{\prime}$ are cached are similar.
Since $D(t,\mathcal{C})$ is the un-normalized delay, we have
\begin{equation}\label{eqn_Proposition1-5}
D^{\prime}(t,\mathcal{C}^{\prime})  :  D^{\prime}(t,\mathcal{C}) \, = \,  |\mathcal{C}^\prime| : |\mathcal{C}| =x.
\end{equation}

Based on the expressions above, we can deduce that
\begin{eqnarray*}
\begin{array}{lll}
D^{\,\prime}(t) & = & \dfrac{1}{\phi^{t\,\,\,\,\,\,\prime}_{sum}U^t_{sum}} \cdot \left[ D^{\prime}(t,\mathcal{C}) + D^{\prime}(t,\mathcal{C^{\prime}}) \right] = \dfrac{1}{\phi^{t\,\,\,\,\,\,\prime}_{sum}U^t_{sum}} \cdot (1+x) \cdot D^{\prime}(t,\mathcal{C}) \\
 & < & \dfrac{1}{\phi^{t\,\,\,\,\,\,\prime}_{sum}U^t_{sum}} \cdot (1+x) \cdot D(t,\mathcal{C}) = \dfrac{1}{\phi^t_{sum}U^t_{sum}} \cdot D(t,\mathcal{C}) = D(t).
\end{array}
\end{eqnarray*}
This result can also be intuitively comprehended that the increase in content number leads to the decrease in caching percentage, making the caching system less efficient.
\end{IEEEproof}

\vspace{-3mm}
\begin{proposition}
With fixed distributions of users, SBSs and content popularity, supposing that the storage capacities of SBSs are the same, then average storage capacity $H$ has a negative correlation with average delay $D(t)$.
\end{proposition}
\vspace{-3mm}
\begin{IEEEproof}
Assume that the average delay with storage capacity $H$ is $D(t)$, and the average delay with storage capacity $H^\prime$ is $D^{\,\prime}(t)$, where $H^\prime > H$.
Since the caching result of $H^\prime$ can be derived from the given caching result of $H$, additional contents can be added to the caching storages, which directly makes $\theta^{t,j \,\,\prime}_{l,k} \le \theta^{t,j}_{l,k}$.
Note that at least one set of $l,k$ leads $\theta^{t,j \,\,\prime}_{l,k} < \theta^{t,j}_{l,k}$, thus we have $D^{\,\prime}(t)<D(t)$.
\end{IEEEproof}

\begin{figure}[!thp]
\centering
\includegraphics[width=4.5in]{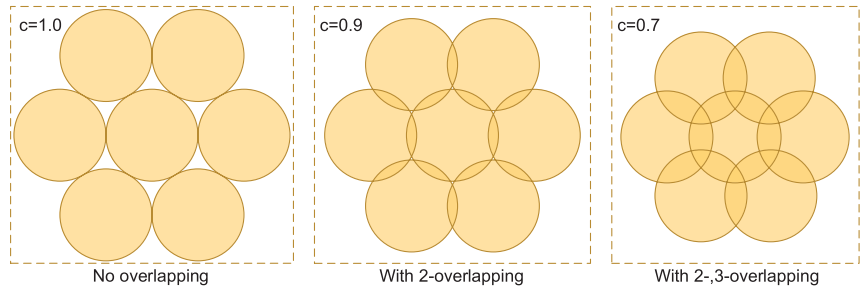}
\vspace{-6mm}
\caption{A top view of the overlapping regions of SBSs with radius $R$. The distance of two adjacent SBSs is $2Rc$ where $c$ is the compress factor. We let $c \in [1/\sqrt{3},1]$ to make sure that overlapping regions of four or more SBSs don't exist.}\label{fig_Intersection}
\vspace{-6mm}
\end{figure}

Unlike the analysis on number of content or storage capacity, the influence of overlapping is abstruse due to the complicated geographic distribution of SBSs.
We have to first assume that SBSs with fixed coverage radius are uniformly distributed in a cellular grid, where we control the overlapping percentage by making the cellular grid denser or sparser.
An illustration is shown in Fig.~\ref{fig_Intersection}.
To further simplify the problem to be analyzed, we only consider a special case where the parameter $\beta_3$ in the equation (\ref{eqn_DelayChoose}) equals to zero, i.e., the choosing delay is ignored.

\vspace{-3mm}
\begin{proposition}
In an approximately infinite cellular grid where SBSs with fixed coverage radius are uniformly distributed, given the constraint that 1) users density is uniform, 2) no coverage regions of four or more SBSs exist, and 3) the choosing delay can be ignored, the average delay based on a fixed caching result decreases when the overlapping percentage increases.
\end{proposition}
\vspace{-3mm}
\begin{IEEEproof}
The detailed proof of this proposition is given in the Appendix.
\end{IEEEproof}

These three propositions can be verified by our simulation results provided in section \ref{sec_Simulation}.
From the theoretical analysis above, we can have a rough idea of how well the caching can benefit the system under different circumstances.
Although the total number of contents are not under control in a real world, we can still achieve a better result by enlarging the storage capacity of SBSs.
Besides, the performance can be improved if the overlapping percentage of SBSs is greater, as long as the choosing delay is ignorable.

\vspace{-2mm}
\section{Auction Mechanism Design}\label{sec_Auction}

In this section, we propose an auction-based mechanism to solve the caching problem.
In this mechanism, the caching scheme for each hour is determined by a series of \emph{multi-object auctions}, where the objects are the storages of SBSs that owned by MNOs, and the bidders are the SPs who possess different sets of contents.
Since the contents have different sizes, it is difficult to apply auctions directly based on the original contents.
Therefore, MNOs should announce a \emph{standard caching size $S$} in the auctions.
With this standard, the storages of SBSs are divided into multiple blocks with size of $S$, and the contents of each SP are transformed into $S$-sized independent content blocks.
In addition, we also propose \emph{additional prices} are charged to properly reduce the frequency of content replacement between hours.

For the rest of this section, we first provide the setup of multi-object auctions at each hour, then introduce the market matching algorithm for each auction, and finally we discuss some properties of our mechanism.
The whole procedure of our mechanism is shown in Algorithm \ref{algo_Overview}.

\vspace{-3.5mm}
\begin{algorithm}[!thp]
\caption{The proposed auction-based caching mechanism.}\label{algo_Overview}
\vspace{-1.5mm}
\Begin
{\vspace{-1.5mm}MNOs announce the standard caching size $S$\;
\vspace{-1.5mm}Auctions setup at $t=1$\;\vspace{-1.5mm}
\While{true}
{\vspace{-1.5mm}Each SP transforms its own contents to form multiple $S$-sized content blocks\;\vspace{-1.5mm}
    \For{$j$ is from $1$ to $\max\limits_i\{H_{i}/S\}$}
     {\vspace{-1.5mm}The $j^{th}$ storage blocks in all SBSs are regarded as objects\;
     \vspace{-1.5mm}SPs estimate the utility of caching each of their content blocks to each of SBSs\;
     \vspace{-1.5mm}Create the valuation matrix based on current allocations and additional prices\;
     \vspace{-1.5mm}Run the \emph{market matching algorithm} to complete one single multi-object auction\;\vspace{-1.5mm}}
     \vspace{-1.5mm}Let $t=t+1$, continue to determine the caching result in the next hour\;\vspace{-1.5mm}
}\vspace{-1.5mm}
}\vspace{-1.5mm}
\end{algorithm}

\vspace{-3mm}
\subsection{Multi-object Auction Setup}\label{sec_Auction_Mechanism}

In this subsection, we first provide a method of transforming contents into equal-sized content blocks, then demonstrate the auctions at each hour and the valuations without additional prices, and finally take addition prices into account and provide the final valuations.

\subsubsection{Transforming contents into equal-sized content blocks}
\emph{}

\begin{figure}[!thp]
\centering
\includegraphics[width=4in]{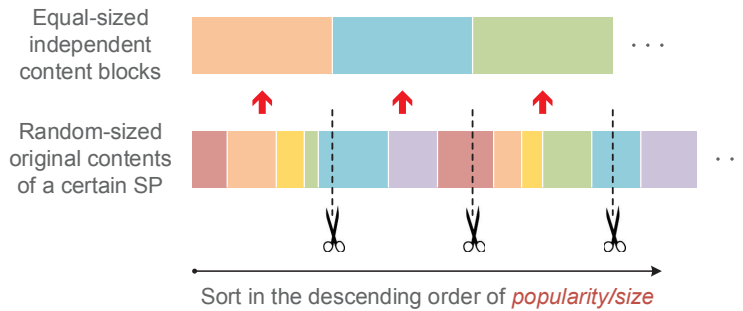}
\vspace{-6mm}
\caption{The method of transforming random-sized original contents of a certain SP into equal-sized independent content blocks.}\label{fig_ContentBlock}
\vspace{-6mm}
\end{figure}

Since the caching problem we've formulated is similar to the classical knapsack problem~\cite{bib_Knapsack}, the transforming procedure that we propose is inspired by one of the greedy algorithm.
For a given SP, we sort all of its contents in the descending order of \emph{popularity to size ratio}, and put them together to form a one-dimensional long ``data ribbon", as shown in Fig.~\ref{fig_ContentBlock}.
And based on the given standard caching size $S$, we cut this ``data ribbon" from the left side into multiple $S$-sized content blocks.
Here we ignore the minor problem that whether the length of the ``data ribbon" can be divisible by $S$, since the most right side usually consists of low-popularity contents and they have little impact to the caching performance.

\vspace{-6mm}

Due to the huge number of contents in reality, we recommend that $S$ is set greater than the largest original content, in which way the computational complexity can be reduced to some extent.
And as a result, each content is divided into no more than two content blocks.
Notice that the newly formed content blocks are independent from each other, which means that no two content blocks share a common slice of data.

Here, we use $B^t_{l,r}$ to denote the $r^{th}$ content block of SP$_l$ at the $t^{th}$ hour.
And the popularity of $B^t_{l,r}$ can be calculated as $\varphi^t_{l,r} = \sum\limits_k \eta^t_{l,r,k} \cdot \phi^t_{l,k}$, where $0 \le \eta^t_{l,r,k} \le 1$, representing the percentage that the original content $C_{l,k}$ is contained in the content block $B^t_{l,r}$.
This equation is essentially to linearly add up the weighted popularity of the contained contents, which conforms to the linear additive formulation given in equation (\ref{eqn_SpecificDelay}).

In the multi-object auctions, we only consider the content blocks as the whole caching objects, and we use $\gamma^{t,i}_{l,r}$ to represent the allocation matrix of content blocks.
Note that this denotation differs from $\gamma^{t,i}_{l,k,n}$ in equation (\ref{eqn_Allocation}), which stands for the allocation of original contents.

\subsubsection{A serial of multi-object auctions for each hour}
\emph{}

\begin{figure}[!thp]
\centering
\includegraphics[width=3.8in]{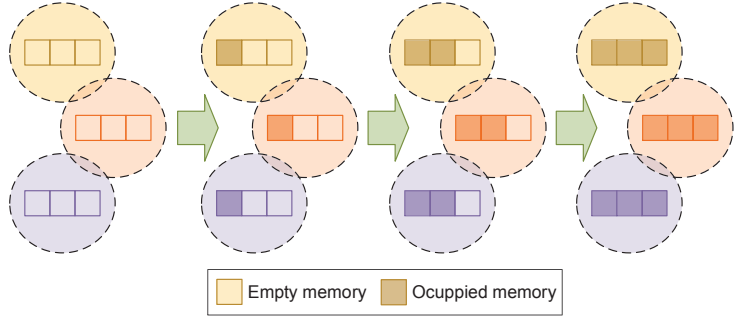}
\vspace{-6mm}
\caption{The mechanism for each hour's caching, where a dashed circle indicates the coverage region of a SBS.}\label{fig_Auction}
\vspace{-6mm}
\end{figure}

The caching problem for each hour is solved by holding a serial of multi-object auctions.
Specifically, we auction for $\max\{\dfrac{H_i}{S}\}$ times, where the $j^{th}$ memory block in all the SBSs are auctioned off in the $j^{th}$ auction.
This process is essentially to auction the storages of all SBSs concurrently with multiple steps, as shown in Fig.~\ref{fig_Auction}.
In each multi-object auction, SPs play the roles of bidders and storages of SBSs play the roles of objects.
After the each auction, each SP obtains a certain amount of caching spaces in each SBSs.
Then each SP can place its contents into SBSs according to the caching result (can be done automatically by its server).

\vspace{-3mm}

However, before each auction, SPs have to estimate the utility of caching each of their content blocks to each of the SBSs and bid for them.
Based on equations (\ref{eqn_SpecificDelay}) and (\ref{eqn_AverageDelay}), we give the expression to calculate the marginal utility of caching each content block into each SBS during the auction procedure as:
\begin{equation}\label{eqn_Value}
V_{l,r}^{t,i} = \sum\limits_{j|i\in A_j} -\Delta \theta^{t,j}_{l,r} \cdot U^t_j \cdot \varphi^t_{l,r},
\end{equation}
where $-\Delta \theta^{t,j}_{l,r}$ is the average decrease of latency for the users in $\Omega_j$ requesting contents in $B^t_{l,r}$, if $B^t_{l,r}$ is newly added into the storage of SBS$_i$ during the procedure of allocation.

\subsubsection{Additional prices for content replacement between hours}
\emph{}

Since content popularity and user density are time variant, caching results in different hours may differ a lot.
When a SBS changes its caching contents, additional instantaneous traffic load burdens the backhaul.
Regardless the specific technique to change or pre-cache contents, we define $\lambda^t$ as the \emph{replacement percentage} at the $t^{th}$ hour to indicate this additional load, given by

\vspace{-3mm}
\begin{equation}
\lambda^t = \dfrac{1}{I} \sum\limits^{I}_{i=1} \dfrac{1}{B_i} \sum\limits_{l,r} S \cdot (1-\varepsilon^{t,i}_{l,r}) \cdot \gamma^{t,i}_{l,r} ,
\end{equation}
where $\varepsilon^{t,i}_{l,r}$ represents the percentage of data in $B^t_{l,r}$ that was cached in SBS$_i$ at $t-1$.
Therefore, $\lambda^t$ shows the average percentage of the storages of SBSs that are replaced.

In order to reduce $\lambda^t$, additional prices are charged for replacing the original contents if the new ones weren't cached in this SBS in the last hour.
Here, we design the additional price as $\Delta p^{t,i}_{l,r}=\omega \cdot (1-\varepsilon^{t,i}_{l,r}) \cdot \theta^t_{back}$, which indicates heavier traffic needs higher additional prices to limit the replacement percentage.
And the constant $\omega$ is defined as \emph{additional price coefficient}.

The introduction of additional price results in an adjustment to the valuations given above.
For $B^t_{l,r}$, the valuation is calculated by $V_{l,r}^{t,i}-\Delta p^{t,i}_{l,r}$.
This is because once $B^t_{l,r}$ obtains the caching storage in SBS$_i$, it will lose another $\Delta p^{t,i}_{l,r}$ because of additional price.

To simplify the denotations later in this section, we use $C_n$ to represent the $n^{th}$ content block (among all the content blocks from all the SPs), where $1\le n \le N$, and use $D_m$ to represent the $m^{th}$ storage block (i.e., the storage block that provided byx SBS$_m$), where $1 \le m \le M$.
The valuation of $C_n$ for $D_m$ is denoted by $v_{n,m}$, which can be calculated by the corresponding expression of $V_{l,r}^{t,i}$ and $\Delta p^{t,i}_{l,r}$, given by $ v_{n,m} = V_{l,r}^{t,i}-\Delta p^{t,i}_{l,r}$.

\vspace{-5mm}
\subsection{Market Matching Algorithm}\label{sec_Auction_Algorithm}

To solve a single multi-object auction, we provide the \emph{market matching algorithm}, which is originated from \cite{bib_MultipleItems} and is able to match the content blocks and the storage blocks with maximum total utility\footnote{The original market matching algorithm was proved to satisfy the VCG principle \cite{bib_ElicitionOfHonest}, where bidders' best strategy is to bid truthfully. However, in our situation, where we regard each content as the corresponding bidder in the original algorithm, the utility of each SP may not be maximized by truthful biding. Different content blocks may belong to the same SP, who aims to maximize its overall utility. Therefore, SP's best strategy may not be truth-telling. But in reality, SPs do not know the valuations of others, so truth-telling still remains a good strategy, because the average utility is excepted to decrease if a SP randomly vary its bids from the true valuations. Therefore, it is reasonable for us to use valuations as the bids in our algorithm.}.
For writing simplicity, we use the word ``content" instead of ``content block" and use ``storage" instead of ``storage block" in the rest of this subsection.

To be brief, this algorithm takes the valuations as the input, uses bipartite graph to get a perfect matching between contents and storages\footnote{Since the algorithm is based on matching, one content cannot get more than 1 storage in each auction. Therefore a problem may arise if $I > \min\{{H_i}\}$, i.e., a content with great popularity is unable to be cached in every SBSs. Hence, we assume $I \le \min\{{H_i}\}$, which can be satisfied in most real-world situations because the storages are usually large enough.}, and outputs the allocation results and the prices of storages.
This algorithm can be described by 7 steps as follows.
Step 1 and Step 2 introduce the initialization process, Step 3 to Step 6 provide the iteration process to find a perfect matching between contents and storages, and Step 7 provides the outcome of the algorithm.

\textbf{Step 1:}
Given $N$ contents and $M$ storages ($N>M$), add $N-M$ virtual storages.

This step is to equalize the number of contents and the number of storages, which is a necessary condition for the following steps.
Due to the unworthiness of virtual storages, the valuations of them are confined as zero and the contents that obtains a virtual objects actually obtains nothing.
And for clearer writing, we assume $N=M$ later in the algorithm introduction.

\textbf{Step 2:}
Initialize the prices of all storages as zero, i.e., $p_m=0$, $\forall m \in [1, M]$.

The price of a storage represents the money that has to be paid by the SP whose content obtains this storage.
And these prices will gradually increase in the process of the algorithm.

\begin{definition}
\vspace{-3mm}
In a bipartite graph where a set of content nodes $\mathcal{C}$ is connected to a set of storage nodes $\mathcal{D}$, the edge between $C_n$ and $D_m$ exists if and only if $(v_{n,m}-p_m)$ is the largest for any $m$ with a fixed $n$, then the bipartite graph is called a \emph{preferred-storage graph}.
\vspace{-3mm}
\end{definition}

\textbf{Step 3:}
Based on the valuations and the prices, a preferred-storage graph can be built.
To put it simple, the preferred-storage graph shows which are the most preferred storages of each content.
For $C_n$, if the profit of acquiring $D_m$ is highest, then there will be an edge between $C_n$ and $D_m$.
Note that for each content, there may be serval most preferred storages.

\begin{definition}
\vspace{-3mm}
Given a graph $G=(V,E)$, a \emph{matching} is a subset edges of $E$ such that no two edges in this subset share a same vertex.
\end{definition}

\begin{definition}
\vspace{-3mm}
Given a graph and a matching of it, an \emph{alternating path} is a serial of consecutively connected edges such that these edges are alternately contained or not contained in the matching.
And an alternating path is an \emph{augmenting path} if and only if the two end-vertices in the alternating path are unmatched.
\vspace{-3mm}
\end{definition}

\textbf{Step 4:}
In the preferred-storage graph, we use augmenting paths to expand the matching until no augmenting paths can be found.

The classical breadth-first-search (BFS) algorithm \cite{bib_GraphIntro} is applied to find augmenting paths from any of an unmatched content node.
Given a matching $\mathcal{M}$ and a certain augmenting path, we denoted all the edges in the augmenting path as $\mathcal{E}$.
The edges both in $\mathcal{E}$ and $\mathcal{M}$ are denoted as $\mathcal{E}_1$, and the edges in $\mathcal{E}$ but not in $\mathcal{M}$ are denoted as $\mathcal{E}_2$.
By adding $\mathcal{E}_2$ to the matching and deleting $\mathcal{E}_1$ from the matching, a greater matching can be formed, because $|\mathcal{E}_2| = |\mathcal{E}_1| +1$.
The matching achieves maximum when there are no augmenting paths can be found.

\textbf{Step 5:}
Based on the matching in the last step, if all the nodes are matched, i.e., the maximum matching is a perfect matching, then jump to the step 7.
Otherwise, a constricted set can be found, which forbids us to get a perfect matching.
The constricted set is defined as below:

\begin{definition}
\vspace{-3mm}
In a preferred-storage graph, given $\mathcal{C}^\prime$ as a subset of the content nodes, denote the directly connected storage nodes as set $\mathcal{D}^\prime$.
If $|\mathcal{C}^\prime| > |\mathcal{D}^\prime|$, then $\{ \mathcal{C}^\prime , \mathcal{D}^\prime \}$ forms a \emph{constricted set}.
\vspace{-8mm}
\end{definition}

\begin{figure}[!thp]
\centering
\includegraphics[width=3.5in]{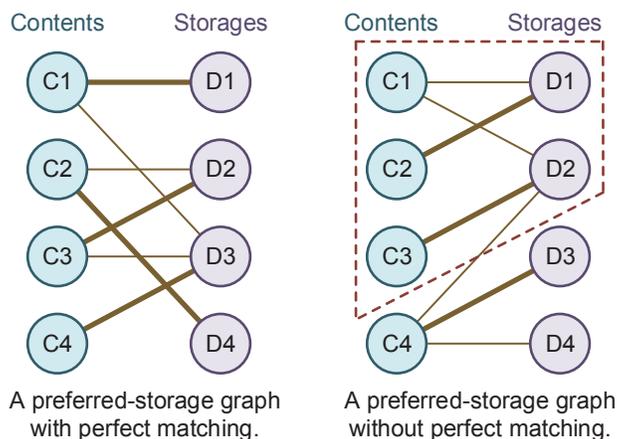}
\vspace{-8mm}
\caption{Two preferred-storage graphs. The left one has a perfect matching, while the right one is confined by a constricted set shown in the dashed box.}\label{fig_Bipartite}
\vspace{-6mm}

\end{figure}

Intuitively, the constricted set cannot form a perfect matching in itself because the number of storages is larger.
Therefore, the whole bipartite graph fails to form a perfect matching if a constricted set is contained.
\cite{bib_MatchingTheory} shows that the equivalence condition of a bipartite graph having a perfect matching is that there are no constricted sets.
The process of searching for a constricted set is simple.
When the algorithm fails to find an augmenting path during BFS, the nodes that being visited by BFS automatically form a constricted set \cite{bib_MatchingTheory}.
Two examples of perfect matching and constricted set are shown in Fig.~\ref{fig_Bipartite}.

\textbf{Step 6:}
Once a constricted set $\{\mathcal{C}^\prime , \mathcal{D}^\prime\}$ is found, the algorithm raises the prices of storages in $\mathcal{D}^\prime$ uniformly, until at least one content changes its preferred-storages so that a new preferred-storage graph can be built.
If $\min\{p_m =\delta >0\}$, let $p_m=p_m-\delta$ for all $m$.
This step is necessary to keep the price as the lowest market-clearing price to obey the VCG principle \cite{bib_MultipleItems}.
After that, the algorithm goes back to step 3 and continues to build a new preferred-storage graph.

\textbf{Step 7:}
The algorithm ends.
The matching shows the allocation between contents and storages.

As a summary, an overview of the whole algorithm is given in Algorithm \ref{algo_Market}.

\vspace{-1.5mm}
\begin{algorithm}[!thp]
\caption{Market matching algorithm for each multi-object auction.}\label{algo_Market}
\vspace{1.5mm}
\KwIn{Valuation matrix $V_{N\times M}$ ($N$: contents, $M$: storages, $N>M$).}\vspace{-1.5mm}
\KwOut{Allocation matrix $\Gamma_{N\times M}$ and price vector $P_{M}$.}\vspace{-1.5mm}
\Begin
{\vspace{-1.5mm}
Expand the valuation matrix from $V_{N\times M}$ to $V_{N\times N}$ with zeros (add virtual storages)\;\vspace{-1.5mm}
Initialize the price vector $P_{N}$ as zero\;\vspace{-1.5mm}
\While{true}
{\vspace{-1.5mm}
    Build a preferred-storage graph $G(\mathcal{C},\mathcal{D},\mathcal{E})$ based on $V_{N\times N}$ and $P_{N}$\;\vspace{-1.5mm}
    Find a maximum matching ${M}$ in this preferred-storage graph\;\vspace{-1.5mm}
    \eIf{${M}$ is a perfect matching}
    {\vspace{-1.5mm}Break the while loop\;\vspace{-1.5mm}}
    {\vspace{-1.5mm}
    Find a constricted set $(\mathcal{C}^\prime , \mathcal{D}^\prime)$ in $G(\mathcal{C},\mathcal{D},\mathcal{E})$\;\vspace{-1.5mm}
    Find the minimum price $\delta p$, which can change $G$ once added to storages in $\mathcal{D}^\prime$\;\vspace{-1.5mm}
    Let $p_n=p_n+\delta p$ for all $n$ that satisfies $D_n\in \mathcal{D^\prime}$\;\vspace{-1.5mm}
    Let $p_n=p_n-\min{\{ p_n \}}$ for all $1\le n\le N$\;\vspace{-1.5mm}
    }\vspace{-1.5mm}
}\vspace{-1.5mm}
The allocation matrix $\Gamma_{N\times M}$ shows the caching result\;\vspace{-1.5mm}
}
\end{algorithm}

\vspace{-5mm}
\subsection{Properties of the mechanism}\label{sec_Auction_Mechanism}

In this subsection, we first discuss the influence of additional prices, then prove the convergence of the algorithm, and finally calculate its complexity.

\subsubsection{Influence of Additional Prices}
\emph{ }

The proposal of additional prices can limit the frequency of content replacement, especially at busy hours.
Although the advantage of setting additional prices is non-trivial as shown in section \ref{sec_Simulation}, the disadvantage still exists.
When $\omega$ is too high, the caching result may not change between adjacent hours, which leads to an inefficient performance due to the time-dependent content popularity and user density.
Therefore, choosing proper $\omega$ for the system is actually a tradeoff between the load of content replacement and the effectiveness of caching in each hour.

\subsubsection{Convergence of the Market Matching Algorithm}
\emph{ }

\vspace{-3mm}
\begin{proposition}
Given that the valuations are presented by decimals with finite precision and finite upper-bound, the algorithm has convergence.
\end{proposition}
\vspace{-3mm}
\begin{IEEEproof}
We define \emph{the content's potential profit}, $P_c^n$ as the maximum profit that content $C_n$ can currently obtain from any one of the storages, and define \emph{the storage's potential profit} $P_s^m$ as the price of the storage $D_m$.
The sum of all the potential profit of contents and storages $P_{sum}$ represents the current maximum possible social welfare.
Note that the existence of constricted set $\{\mathcal{C^\prime , D^\prime}\}$ makes it unable to satisfy all the $C_n \in \mathcal{C}^\prime$ obtaining their profits, so $P_{sum}$ may be exaggerated.
At the beginning of the algorithm, $P_{sum}^{beg}\ge 0$ because $P_c^n \ge 0$ and $P_s^m \ge 0$ for any $n$ and $m$.
And at the end of the algorithm, we have $0 \le P_{sum}^{end}\le P_{sum}^{beg}$ because a possible social welfare is lower than an exaggerated one.
In the algorithm, once the minimum price is above zero, we reduce the prices of all storages.
This step doesn't change $P_{sum}$ because $N=M$ and the total decrease of $P_s^m$ equals to that of $P_c^m$.
But when to raise the prices of storages in a constricted set $\{\mathcal{C^\prime , D^\prime}\}$, $P_{sum}$ decreases by $\Delta P$ because $ | \mathcal{C}^\prime | > | \mathcal{D}^\prime | $.
Since $\Delta P >0$ and $\le P_{sum}^{beg} - P_{sum}^{end}$ is finite, $P_{sum}$ can finally decrease to $P_{sum}^{end}$ after certain amounts of iterations.
\end{IEEEproof}

\subsubsection{Complexity of the Market Matching Algorithm}
\emph{ }

The proof of its convergence shows that the complexity of this algorithm depends on the precision of valuations.
We define the quantification accuracy as $\alpha =  V / D$, where $V$ is the maximum possible valuation  and $D$ is the minimum division of valuations.
For the case that maximum possible valuation is 100 and minimum division of value is 0.1, we have $\alpha=1000$.

\vspace{-3mm}
\begin{proposition}
Given $N$ as the total number of content blocks, the complexity of the market matching algorithm is $O(\alpha N^4 )$.
\end{proposition}
\vspace{-3mm}
\begin{IEEEproof}
The algorithm requires no more than $\alpha \cdot N$ times of iterations to get a perfect matching, since $0\le P_{sum}^{end} \le P_{sum}^{beg} \le VN$ and $\Delta P \ge D$.
At the beginning of each iteration, a preferred-storage graph is constructed in the complexity of $O(N^2)$.
Then, less than $N$ times of BFS are executed, which takes $O(N^3)$.
Finally, a constricted set is found and prices are changed in $O(N^2)$.
Therefore, the algorithm takes $O(N^3)$ in each iteration, implying that the whole algorithm is $O(\alpha N^4 )$.
\end{IEEEproof}

As shown in our simulations, the practical complexity of this algorithm is not as high as $O(\alpha N^4 )$.
Moreover, we can set the standard caching size $S$ greater, to adapt to the enormous number of content.

\section{Simulation Results}\label{sec_Simulation}

In this section, we simulate the performance of the proposed mechanism, the impact of system parameters, and the influence of the quantification accuracy.
The simulation parameters are set in the first subsection, the simulation results and discussions are provided in the second subsection.

\vspace{-5mm}
\subsection{Simulation Parameters}

Without loss of generality, we set $H_i= H$ for all $1 \le i \le I$.
According to \cite{bib_LoadProfile}, the traffic loads in different days have the similar profile, so we set the variation of average user density $u(t)$ in a similar way, as shown in Table \ref{tab_Parameters}.
At each hour, the user density of each region conforms to Poisson distribution with mean value of $u(t)$.

\begin{table}[!thp]
\renewcommand\arraystretch{0.9}
\caption{Simulation parameters}\label{tab_Parameters} \centering
\vspace{-2mm}
\begin{tabular}{|p{45mm}|p{80mm}|}
\hline
Popularity parameter $\mu$ and $\sigma$ & $1$ and $0.5$ \\
\hline
Popularity parameter $a$, $b$ and $t_0$ & $\mathcal{U} [0, 3]$, $\mathcal{U} [4, 12]$, and $\mathcal{U} [-75, 25]$ \\
\hline
Average user density $u(t)$ in 24 hours respectively ($10^{-5} m^{-2}$) & $380$,$210$,$110$,$110$,$140$,$200$,$300$,$650$,$1100$,$1260$,$1400$,$1570$,   $1530$,$1370$,$1310$,$1250$,$900$,$800$,$940$,$1100$,$1200$,$1070$,$610$,$450$\\
\hline
Delay parameter $\beta_1$, $\beta_2$, $\beta_3$ (ms) & $1$, ~$5$, ~from $0$ to $400$\\
\hline
Number of SBSs $I$ & $24$ \\
\hline
Radius of SBSs $R$ (m) & $50$\\
\hline
Total content number $K$ & from $10000$ to $20000$ \\
\hline
Overlapping percentage $O$ & from $20\%$ to $100\%$ \\
\hline
Size of contents $S_{l,k}$ (GB) & $\mathcal{U} [0.1, 1]$ \\
\hline
Average storage capacity $H$ (GB) & from $0$ to $1000$ \\
\hline
Size of content blocks $S$ (GB) & $20$ \\
\hline
Additional price coefficient $\omega$ & from $0$ to $4$ \\
\hline
Quantification accuracy  $\alpha$ & from $10$ to $1000$ \\
\hline
\end{tabular}
\vspace{-5mm}
\end{table}

Since the popularity of $C_{l,kx}$ is time-dependent, we assume that it has a similar time-evolutionary profile with the log-normal probability density function.
This assumption accords with the study of \cite{bib_Exp} in characterizing the slow fading popularity of contents from time domain.
And it also guarantees that the popularity distribution of large amount of contents at any given time conforms to Zipf-like distribution \cite{bib_Zipf}.
The log-normal probability density function is given by
\begin{equation}
f(x) = \left\{
            \begin{array}{lcc}
            \frac{1}{\sqrt{2 \pi} \sigma x} \exp{\left[ \frac{-(\ln{x}-\mu)^2}{2 \sigma^2} \right]}, & &  x> 0, \\
            0,& & x \leq 0,
            \end{array}
            \right.
\end{equation}
where the parameter $\mu$ and $\sigma$ can be properly selected.
Note that the maximum popularity and lifespan of different contents can be distinct from each other, therefore, we add extra parameters to the original function as $\phi_{l,k}^t = a f(\frac{t-t_0}{b}) $,
where $a$ determines the maximum popularity, $b$ represents the lifespan, and $t_0$ is the time when $C_{l,k}$ is uploaded.

\begin{figure}[!thp]
\centering
\includegraphics[width=4.5in]{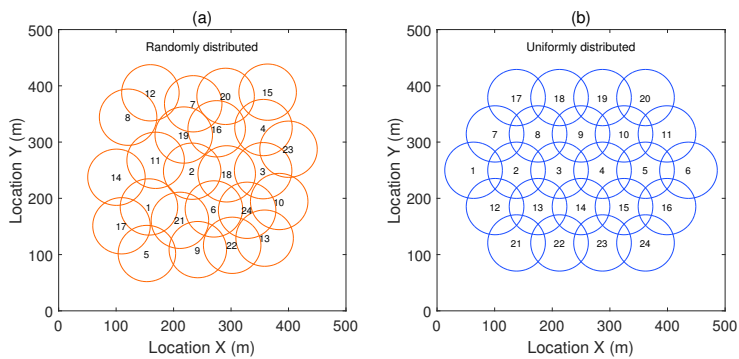}
\vspace{-6mm}
\caption{Two SBSs-distribution settings. The left figure shows an example of the uniformly randomly generated distribution of 20 SBSs. The right figure shows a fixed uniform distribution of 24 SBSs.}\label{fig_SBS}
\vspace{-8mm}
\end{figure}

The SBSs are set in two different ways, as shown in Fig.~\ref{fig_SBS}, the random distribution and the uniform distribution.
Here, we make sure that no more than three SBSs overlaps with each other, which is both for reality and simplicity.\footnote{The goal of deploying SBSs is to provide higher data transmission rate and larger coverage area, so it is unwise to put too many SBSs within a small region. In addition, the property we discussed in Proposition~3 intuitively has the same trend when we extend the area of overlapping from 3 to four or more, thus this constraint does not lose any generality.}

Table \ref{tab_Parameters} shows a more detailed list of parameters, where $\mathcal{U} [P,Q]$ means that the probability density is uniformly distributed from $P$ to $Q$.

\vspace{-5mm}
\subsection{Simulation Results and Discussions}

\begin{figure}[!thp]
\centering
\includegraphics[width=3in]{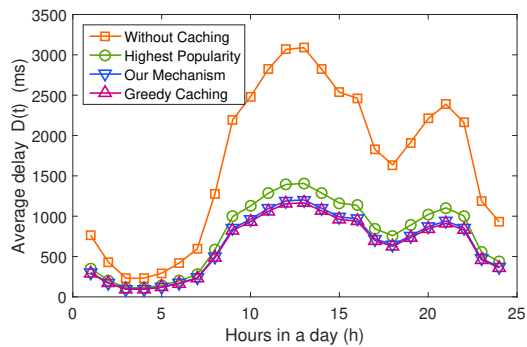}
\vspace{-6mm}
\caption{Average delay profile in one day, where $O=54\%$, $K=10000$, $H=1000 GB$, $\beta_3=0 ms$ and $\omega=0$.}\label{fig_OneDay}
\vspace{-6mm}
\end{figure}

We first simulate a one-day situation to show how the average delay changes in a day, as given in Fig.~\ref{fig_OneDay}, where $O=54\%$, $K=10000$, $H=1000 GB$, $\beta_3=0 ms$, and $\omega=0$.
The uppermost curve shows how the average delay changes during a day without caching according to the predefined values in Table~\ref{tab_Parameters}.
And the other three curves shows the average delay with different caching strategies.
\emph{Highest popularity} means caching the most popular contents in each SBS.
\emph{Greedy caching} comes from the algorithm proposed in \cite{bib_RelatedFemtoCaching}, which allocates only one content block in each round (while our algorithm allocates $I$ content blocks in each round).
The outcome of our mechanism and the greedy caching algorithm are quite similar, and both surpass the highest popularity algorithm.
Note that, among these caching schemes, only our mechanism is designed to solve the problems of multiple SPs.
The reason we put them in the same figure is to show the effectiveness of our mechanism by comparing it with the existing caching schemes.

\vspace{-6mm}

\begin{figure}[!thp]
\centering
\includegraphics[width=3in]{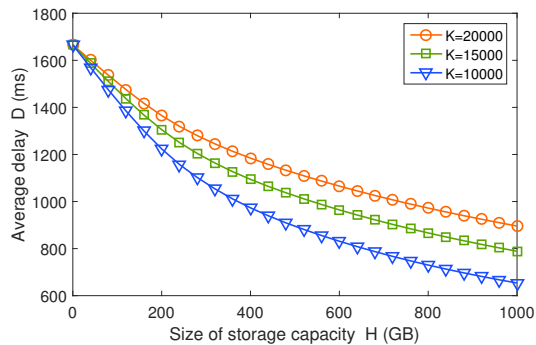}
\vspace{-6mm}
\caption{Average delay $D$ verses storage capacity $H$, with $K=20000,15000,10000$, respectively.}\label{fig_Simulation_Capacity_ContentNumber}
\vspace{-6mm}
\end{figure}

The impact of $H$ and $K$ is shown in Fig.~\ref{fig_Simulation_Capacity_ContentNumber}, where we set $O=54\%$, $\beta_3=0 ms$, $\omega=0$, and let $K=20000,15000,10000$ respectively.
Here we use $D$ to denote the average delay in 24 hours, given by $D= \dfrac{1}{24} \sum_{t=1}^{24} D(t)$, where $D(t)$ is the average delay of each hour.
In this figure, we can see that when $H$ gets greater, $D$ decreases but the change rate of $D$ decreases as well.
So the same amount of storage makes greater difference in a low-capacity situation.
It can also be observed that a greater number of contents $K$ makes it more difficult to achieve low latency.
Therefore, the simulation on the impact of $K$ and $H$ agrees with Propositions 1 and 2.

\begin{figure}[!thp]
\centering
\includegraphics[width=6.5in]{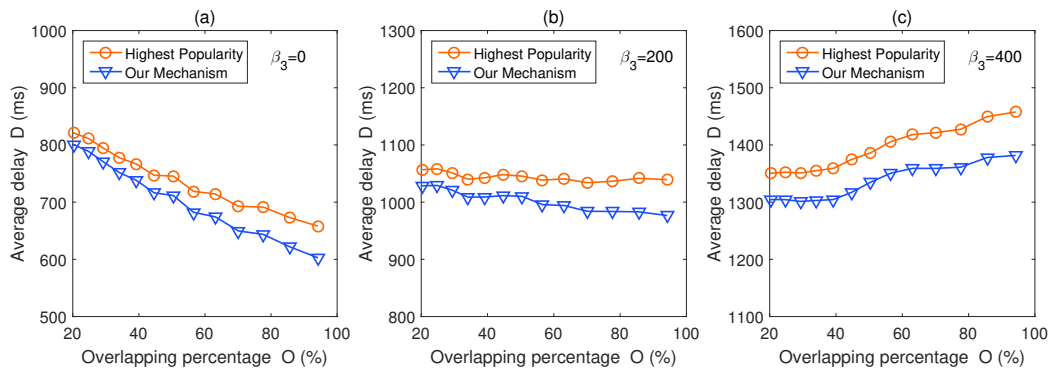}
\vspace{-12mm}
\caption{Average delay $D$ verses overlapping percentage $O$, where $\beta_3=0ms, 200ms, 400ms$, respectively.}\label{fig_Simulation_Overlapping}
\vspace{-6mm}
\end{figure}

The impact of $O$ is shown in Fig.~\ref{fig_Simulation_Overlapping}, where $H=1000GB$, $K=10000$, and $\beta_3=0ms, 200ms, 400ms$, respectively.
Here we control $O$ of uniformly distributed SBSs by multiplying the coordinates of SBSs with a constant, which is detailedly described in the Appendix.
From all of the three subplots, we can see that the advantage of our mechanism over \emph{highest popularity} becomes greater if $O$ is higher.
In Fig.~\ref{fig_Simulation_Overlapping}~(a), where the choosing delay $\beta_3$ is ignorable, we find that $D$ decreases with $O$, which agrees with Proposition 3.
However, in Fig.~\ref{fig_Simulation_Overlapping}~(b), where $\beta_3$ is set as $200ms$, the curve of $D$ is becomes flat.
And finally in Fig.~\ref{fig_Simulation_Overlapping}~(c), where $\beta_3=400ms$, the correlation of $O$ and $D$ changes to positive instead of negative.
These results imply that, the delay in overlapping regions can make a great difference on the average outcome.
The caching efficiency can be further improved if we can shorten the ``SBS choosing" procedure of mobile users in a practical cellular system.

\vspace{-6mm}

\begin{figure}[!thp]
\centering
\includegraphics[width=3in]{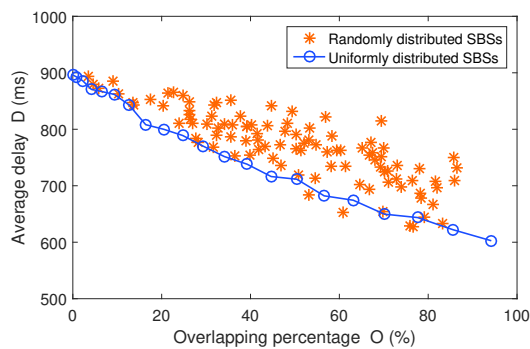}
\vspace{-6mm}
\caption{Average delay $D$ verses overlapping percentage $O$, for uniformly distributed and randomly distributed SBSs.}\label{fig_Simulation_OverlappingRandom}
\vspace{-6mm}
\end{figure}

In Fig.~\ref{fig_Simulation_OverlappingRandom}, we also provide the influence of $O$ in the case of randomly distributed SBSs, with $\beta_3=0$.
Since the number of possible random distributions of SBSs is infinite, and the distributions of SBSs and $O$ have no one-to-one correspondence, we use the uniformly distributed SBSs as the benchmark to observe the outcome of randomly distributed SBSs.
Here, $100$ random cases are generated, and the results are presented by the star points on the figure.
It can be observed that, for a certain $O$, the caching performance is not fixed.
But roughly speaking, the correlation of $O$ and $D$ is similar to that of the uniform distribution, and the correlation coefficient in this simulation is around $-0.8$.
We can also conclude that the floating range of $D$ depends on $O$: A greater $O$ brings $D$ more uncertainty.
And since the line of uniformly distributed SBSs separates most of the star points to the upward side, we can regard the uniform distribution as an effective way to deploy SBSs.

\vspace{-6mm}

\vspace{-6mm}
\begin{figure}[!thp]
\centering
\includegraphics[width=5in]{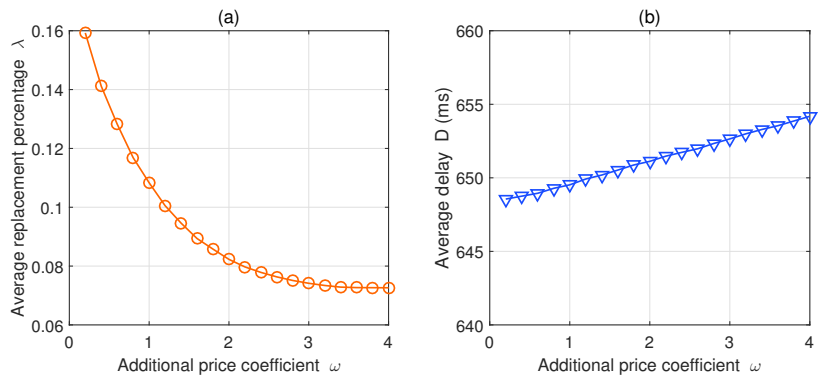}
\vspace{-6mm}
\caption{The influence of additional price coefficient $\omega$ on average replacement percentage $\lambda$ and average delay $D$.}\label{fig_Simulation_AdditionalPrice}
\vspace{-6mm}
\end{figure}

Then, we analyse the impact of $\omega$, where we define $\lambda= \sum_{t=1}^{24} \lambda^t$.
In Fig.~\ref{fig_Simulation_AdditionalPrice}, the relation of $\omega$ and $\lambda$ as well as the relation of $\omega$ and $D$ are given respectively, with $H=1000GB$, $N=10000$ and $O=54\%$.
It indicates that a higher additional price coefficient leads to a lower replacement percentage, but results in a higher average delay.
When $\omega<2$, $\lambda$ decreases sharply but $D$ increases slowly.
This implies that a proper choice of $\omega$ can greatly reduce the load brought by content replacement, with only a trivial cost on the average delay.

\begin{figure}[!thp]
\centering
\includegraphics[width=6.5in]{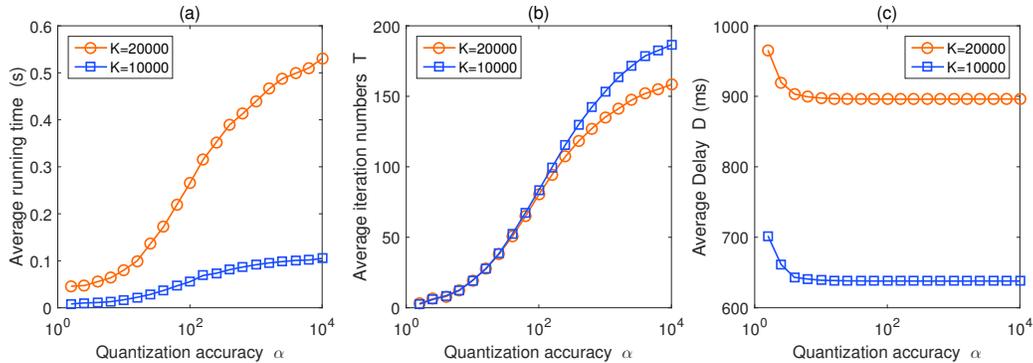}
\vspace{-12mm}
\caption{The influence of quantization accuracy $\alpha$ on average iteration times $\eta$ and average delay ratio $D$ , with $K=10000$ and $K=20000$, respectively, where $O=34\%$, $B=100$.}\label{fig_Simulation_Accuracy}
\vspace{-6mm}
\end{figure}

Finally, we analyse the influence of quantification accuracy $\alpha$, on the time complexity of the algorithm and on the average delay of caching.
Fig.~\ref{fig_Simulation_Accuracy}~(a) shows the average running time (by seconds) of the market matching algorithm to complete each round of auction, where two curves with $K=10000$ and $K=20000$ are given.
And in Fig.~\ref{fig_Simulation_Accuracy}~(b), we count the average iteration number $T$ of this algorithm, i.e., the times of rebuilding preferred-storage graphs to achieve perfect matching.
It can be observed that, $\alpha$ does not contribute to $T$ linearly as the theoretical analysis given by Proposition 5 ( $T=O(\alpha N)$).
What's more, the curve of $K=20000$ is even below the curve of $K=10000$, which indicates that $T=O(\alpha N)$ is an over estimated upper bound.
Thus, the practical complexity of running this algorithm can be far below $O(\alpha N^4)$.
To give another aspect of the impact of $\alpha$, Fig.~\ref{fig_Simulation_Accuracy}.~(c) shows how $\alpha$ influences the system performance.
It can be observed that when $\alpha >10$, the two curves become almost flat and converged to certain values.
Therefore, this algorithm can guarantee its efficiency even the quantification accuracy is not high enough, which can further reduce the practical complexity of executing the algorithm.

\section{Conclusions}

In this paper, we proposed that caching can be applied as a service provided by SPs in mobile networks.
We focused on the small-cell caching scenario and formulate the caching problem as how to minimize the average delay in consideration of the competition among SPs for caching storages.
In the theoretical analysis, we found that average delay has a positive correlation with total content number and a negative correlation with average storage capacity.
In addition, overlapping percentage can also benefit the performance, as long as the choosing delay can be ignored.
To solve the caching problem, we designed a mechanism that based on multi-object auctions, where the convergence of the algorithm can be guaranteed as long as the valuations are presented with finite precision.
Simulation results testified the theoretical analysis and also showed that our solution leads to a better system performance, e.g., the average delay is reduced by $50\%$ when $K=10000$, $H=1000GB$, $O=54\%$.


\begin{appendix}

\textbf{Proposition 3.}
In an approximately infinite cellular grid where SBSs with fixed coverage radius are uniformly distributed, given the constraint that 1) user density is uniform, 2) no coverage regions of four or more SBSs exist, and 3) the choosing delay can be ignored, the average delay based on a fixed caching result decreases when the overlapping percentage increases.

\begin{IEEEproof}
We assume that the coverage radius of a SBS is $R$ and the distance of two adjacent SBSs is $2Rc$, where $c$ is the \emph{compress factor} which has a negative correlation with the overlapping percentage $O$.
A smaller $c$ indicates a smaller cellular grid and results in a greater $O$.
To satisfy the constraint that no coverage regions of four or more SBSs exist, we let $1/\sqrt{3}<c<1$.
In the rest part of this proof, we discuss the influence of $c$ instead of the influence of $O$.

Based on equation (\ref{eqn_DelayBackhaul}) and the assumption that user density is uniform, $\theta^t_{back}$ is only proportional to the area of the total coverage region of SBSs.
When the parameter $c$ gradually decreases from $1$ to $1/\sqrt{3}$, we can deduce that $\theta^t_{back}$ is also decreased.
Therefore, $\theta^t_{back}$ has a positive correlation with $c$, and we rewrite $\theta^t_{back}$ as $\theta^t_{back} (c)$ to express this inter-dependent character, where we have
\begin{equation}\label{eqn_Appendix1}
\dfrac{\partial\theta^t_{back}(c)}{\partial c} >0, \quad\quad 1/\sqrt{3}<c<1.
\end{equation}
The average delay can be seen as the function of $\theta^t_{back} (c)$ and $c$, given by
\begin{equation}\label{eqn_Appendix2}
D(t)=D\big(\theta^t_{back}(c),c\big),  \quad\quad 1/\sqrt{3}<c<1.
\end{equation}
Based on equations (\ref{eqn_SpecificDelay}) and (\ref{eqn_AverageDelay}), we can give that
\begin{equation}\label{eqn_Appendix3}
\dfrac{\partial D\big(\theta^t_{back}(c),c\big)}{\partial \theta^t_{back}(c)} \ge 0,  \quad\quad 1/\sqrt{3}<c<1.
\end{equation}

For any $1/\sqrt{3}<c_1<c_2<1$, we can get $\theta^t_{back}(c_1)<\theta^t_{back}(c_2)$ by using inequality (\ref{eqn_Appendix1}) and get $D\big(\theta^t_{back}(c_1),c_1\big) \le D\big(\theta^t_{back}(c_2),c_1\big)$ by further using inequality (\ref{eqn_Appendix3}).
If we had the condition that $D\big(\theta^t_{back}(c_2),c_1\big) \le D\big(\theta^t_{back}(c_2),c_2\big)$, then the conclusion $D\big(\theta^t_{back}(c_1),c_1\big) \le D\big(\theta^t_{back}(c_2),c_2\big)$ could be obtained and we would have found the monotonicity between $D(t)$ and $c$.
Therefore, in the rest part of this section, we prove that, with constant $\theta^t_{back}$, there is a positive correlation between $c$ and $D(t)$.

Based on equation (\ref{eqn_DelayDownlink}) and the assumption of uniform user density, we can see that $\theta^{t,j}_{down}$ doesn't change with $c$.
Along with the condition of fixed $\theta^t_{back}$ and the condition that choosing delay is zero, based on equation (\ref{eqn_SpecificDelay}), we conclude that the delay of requesting a certain content is a fixed value, independent of the location of users.
Since the caching allocation is also fixed, the only thing that influences $D(t)$ is the coverage percentage of any subset of SBSs in the whole region.
Therefore we focus on how $A_s / A_t$ changes with $c$, where $A_s$ is the area of a subset of SBSs and $A_t$ is the total area of all SBSs.

To simplify the expressions of $A_s$ and $A_t$, we introduce another variable $\theta$, where $c=\cos{\theta}$ and $\theta \in ( 0 , \arccos{\tfrac{1}{\sqrt{3}}} )$.
Note that $\partial c / \partial \theta <0$ in its definition domain.
So our objective is to prove $ \partial (A_s / A_t)  / \partial \theta  > 0 $, which can also be denoted as $f^\prime(\theta)>0$, where $f(\theta)=A_s / A_t$ is the area percentage function.

As the demonstration of the top view shown in the Fig.~\ref{fig_Intersection}, we need to consider two different situations in the definition domain of $\theta$.
The first one only involves 2-overlapping regions where $0<\theta<\tfrac{\pi}{6}$, while the other one involves both 2-overlapping regions and 3-overlapping regions at the same time where $\tfrac{\pi}{6}<\theta<\arccos{\tfrac{1}{\sqrt{3}}}$.

We denote the area of the coverage region of one SBS as $A_1$, the area of 2-overlapping region as $A_2$, and the area of 3-overlapping region as $A_3$.
Their expressions can be simplified as:

\vspace{-5mm}
\begin{equation}
A_1  = \pi R^2, \quad\quad\quad\quad\quad\quad\quad\quad\quad\quad\quad\quad\quad\quad\quad\quad\quad\quad \! \theta\in (0 , \arccos{\dfrac{1}{\sqrt{3}}}),
\end{equation}

\vspace{-7mm}
\begin{equation}
A_2  = 2R^2(\theta - \cos{\theta} \sin{\theta}), \quad\quad\quad\quad\quad\quad\quad\quad\quad\quad\quad \,\,\, \theta\in (0 , \arccos{\dfrac{1}{\sqrt{3}}}),
\end{equation}

\vspace{-5mm}
\begin{equation}
\,\,\,\,A_3  =
\left\{
\begin{array}{ll}
0,  & \quad \theta\in (0 , \dfrac{\pi}{6}), \\
R^2 \big[ 3(\theta-\dfrac{\pi}{6}) + \sqrt{3}\cos^2{\theta}  -3\sin{\theta}\cos{\theta}  \big],  & \quad \theta\in (\dfrac{\pi}{6}, \arccos{\dfrac{1}{\sqrt{3}}}).
\end{array}
\right.
\end{equation}

Now we use $A_1$, $A_2,$ and $A_3$ to express $A_s$ and $A_t$.
For $A_t$ as the total area, we assume that the number of hexagons in the cellular grid is approximately infinite so that the influence of its boundary can be ignored.
To get the proportion of the numbers of $A_1$, $A_2,$ and $A_3$, we find a minimum repeated unit in the infinite grid.
This process is the same as finding a cell in the molecular structure of graphite, leading $A_1:A_2:A_3=1:3:2$.
Thus we have
\begin{equation}
A_t = M_t \cdot (A_1-3A_2+2A_3),
\end{equation}
where $M_t$ is a large integer to represent the total number of SBSs.
In this way, $A_t$ is expressed by its equivalent average coverage area of a single SBS.

For $A_s$ as the area of a subset of SBSs, the proportions of $A_2,$ and $A_3$ in $A_s$ is less than those in $A_t$, because hexagons at the boundary have less overlapping regions.
Hence, we have
\begin{equation}
A_s = M_s \cdot (A_1-xA_2+yA_3),
\end{equation}
where $0\le x<3$, $0\le y<2$ and $M_s$ is the number of the SBSs in the given subset.
Another constraint of $x$ and $y$ is introduced later.

Since $A_3$ has different expressions in different situations, we discuss $\theta\in (0 , \tfrac{\pi}{6})$ and $\theta\in (\tfrac{\pi}{6}, \arccos{\tfrac{1}{\sqrt{3}}})$ respectively.
The former case only involves 2-overlapping regions but the latter case involves 3-overlapping regions as well as 2-overlapping regions, as shown in the Fig.~\ref{fig_Intersection}.

\vspace{3mm}
\textbf{Case 1:}$\quad \theta\in \Big(0 , \dfrac{\pi}{6}\Big).$
\vspace{-3mm}

\begin{equation}
f^\prime(\theta) = \dfrac{M_s}{M_t}\left(\dfrac{A_1-xA_2}{A_1-3A_2}\right)^\prime = \dfrac{M_s}{M_t}\dfrac{(3-x)A_1A_2^\prime }{(A_1-3A_2)^2}.
\end{equation}

Since $A_2^\prime= 2R^2(1-\cos{2\theta})>0$, $3-x>0$ and $A_1-3A_2>0$, we can easily get $f^\prime(\theta)>0$.

\vspace{3mm}
\textbf{Case 2:}$\quad \theta\in \Big(\dfrac{\pi}{6}, \arccos{\dfrac{1}{\sqrt{3}}}\Big).$
\vspace{-3mm}

\begin{equation}\label{eqn_derivation}
f^\prime(\theta) = \dfrac{M_s}{M_t}\left(\dfrac{A_1-xA_2+yA_3}{A_1-3A_2+2A_3}\right)^\prime  =\dfrac{M_s}{M_t}\dfrac{\sqrt{3}\sin{\theta}}{6\cos^3{\theta}}\bigg[(2-y)(\pi-2\theta)+4\theta(1+y-x) \bigg].
\end{equation}

In this case, we need another constraint of $x$ and $y$ to complete the proof.
Given a subset of SBSs, we regard each SBS as a vertex in the planar graph.
Each 2-overlapping region is an edge between two adjacent SBSs, and each 3-overlapping region is a face enclosed by three adjacent 2-overlapping regions.
This abstraction process is illustrated in the Fig.~\ref{fig_Transform}.

\vspace{-3mm}
\begin{figure}[!thp]
\centering
\includegraphics[width=2.0in]{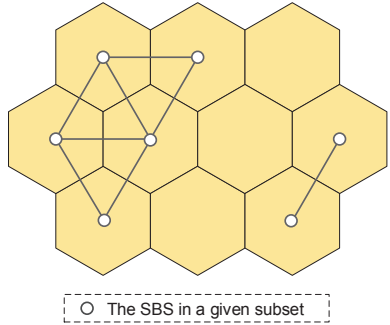}
\vspace{-6mm}
\caption{The transformation from a real situation to a planar graph, where adjacent SBSs in the subset are connected by 2-overlapping regions. The given case indicates that $V=7$, $F=3$, $E=8$ and $C=2$.}\label{fig_Transform}
\vspace{-6mm}
\end{figure}

The Eular Formula in the planar graph \cite{bib_Eular} is given by
\begin{equation}
V+(F+1)-E=C+1,
\end{equation}
where $V$ is the number of vertexes, $F$ is the number of faces, $E$ is the number of edges and $C$ is the number of connected subgraphs.
Thus, the instance in the Fig.~\ref{fig_Transform} implies $V=7$, $F=3$, $E=8$ and $C=2$.
And for any given subset of SBSs, $C\ge 0$ can be satisfied.

We apply the Eular Formula on our model as:
\begin{equation}
(V+F-E)=C\geq 0 \quad \leftrightarrow \quad (V+F-E)/V\geq 0 \quad \leftrightarrow \quad 1+y-x> 0.
\end{equation}

Since $y<2$ and $\theta \in (0,\pi/2)$, based on the equation (\ref{eqn_derivation}), we finally have $f^\prime(\theta)>0$.

\textbf{Conclusion:}
In both cases the area percentage function $f(\theta)$ has a positive correlation with $\theta$ respectively.
When $\theta =\tfrac{\pi}{6}$, $S_3$ remains its continuality, implying that $f(\theta)$ is continuous.
Thus $f(\theta)$ increases with $\theta$ in $(0 , \arccos{\tfrac{1}{\sqrt{3}}})$.
Therefore, the coverage percentage of any given subset of SBSs in the whole area decreases when $c$ gets greater in $(\tfrac{1}{\sqrt{3}},1)$.

As a result, the necessary condition $D\big(\theta^t_{back}(c_2),c_1\big) \le D\big(\theta^t_{back}(c_2),c_2\big)$ mentioned above can be obtained and the positive correlation between $D(t)$ and $c$ can be proved.
\end{IEEEproof}

\end{appendix}

\end{document}